\documentclass[aip,jcp,reprint,eqsecnum,superscriptaddress,twocolumn,longbibliography,floatfix]{revtex4-2}

\usepackage{graphicx}
\usepackage[TABBOTCAP]{subfigure}
\usepackage[utf8]{inputenc}
\usepackage[T1]{fontenc}

\usepackage{mathptmx}
\usepackage{etoolbox}
\usepackage{xcolor}
\usepackage{mathpazo}
\usepackage{bm}
\usepackage[unicode=true,
 bookmarks=true,bookmarksnumbered=false,bookmarksopen=false,
 breaklinks=false,pdfborder={0 0 0},pdfborderstyle={},backref=false,colorlinks=true]
 {hyperref}
\hypersetup{pdfborderstyle={},pdfborderstyle={},pdfborderstyle={},pdfborderstyle={},pdfborderstyle={},pdfborderstyle={},linkcolor=blue,citecolor=blue}
\usepackage{breakurl}

\usepackage{natbib}
\usepackage{amssymb,amsmath}

\usepackage[normalem]{ulem}

\newcommand{\ex}{\text{ex}}
\newcommand{\id}{\text{id}}
\newcommand{\RR}{\text{R}}

\newcommand\beq{\begin{equation}}
\newcommand\eeq{\end{equation}}
\newcommand\beqa{\begin{eqnarray}}
\newcommand\eeqa{\end{eqnarray}}
\newcommand{\nn}{\nonumber\\}
\def\bal#1\eal{\begin{align}#1\end{align}}

\makeatletter
\def\@email#1#2{%
 \endgroup
 \patchcmd{\titleblock@produce}
  {\frontmatter@RRAPformat}
  {\frontmatter@RRAPformat{\produce@RRAP{*#1\href{mailto:#2}{#2}}}\frontmatter@RRAPformat}
  {}{}
}%
\makeatother

\begin{document}

\title[Anisotropic hard bodies in one-dimensional channels]{Ordering properties of anisotropic hard bodies in one-dimensional channels}

\author{Ana M. Montero}
  \email{anamontero@unex.es}
  \affiliation{Departamento de F\'isica,
    Universidad de Extremadura, E-06006 Badajoz, Spain}
\author{Andr\'es Santos}
  \email{andres@unex.es}
  \affiliation{Departamento de F\'isica,
    Universidad de Extremadura, E-06006 Badajoz, Spain}
  \affiliation{Instituto de Computaci\'on Cient\'ifica Avanzada (ICCAEx),
    Universidad de Extremadura, E-06006 Badajoz, Spain}
\author{P\'eter Gurin}
  \email{gurin.peter@mk.uni-pannon.hu}
  \affiliation{Physics Department, Centre for Natural Sciences,
    University of Pannonia, P.O. Box 158, Veszpr\'em H-8201, Hungary}
\author{Szabolcs Varga}
 \email{varga.szabolcs@mk.uni-pannon.hu}
 \affiliation{Physics Department, Centre for Natural Sciences,
   University of Pannonia, P.O. Box 158, Veszpr\'em H-8201, Hungary}

\date{\today}

\begin{abstract}
The phase behavior and structural properties of hard anisotropic particles (prisms and dumbbells) are examined in one-dimensional channels using the  Parsons--Lee (PL) theory, and the transfer-matrix  and  neighbor-distribution  methods. The particles are allowed to move freely along the channel,  while their orientations are constrained such that one particle can occupy only two or three different lengths along the channel. In this confinement setting,   hard prisms behave as an additive mixture,  while  hard dumbbells behave as a non-additive one. We prove that all  methods provide exact results for the phase properties of hard prisms, while only the neighbor-distribution and transfer-matrix methods are exact for hard dumbbells. This shows that non-additive effects are incorrectly included into the PL theory, which is a successful theory of the isotropic-nematic phase transition of rod-like particles in higher dimensions. In the one-dimensional channel, the orientational ordering develops continuously with increasing density, i.e., the system is isotropic only at zero density, while it becomes perfectly ordered at the close-packing density. We show that there is no orientational correlation in the hard prism system, while the hard dumbbells are orientationally correlated with diverging correlation length at close packing. On the other hand, positional correlations are present for all the systems, the associated correlation length diverging at close packing.
\end{abstract}

\maketitle

\section{Introduction}

     It is generally accepted that hard-body interactions play a key
     role in the structural properties and phase behavior of
     molecular liquids, colloids and soft
     matter.\cite{Barrat_2003} In addition to this, the shape of the
     particles is also important, as the anisotropic shape is
     responsible for the stabilization of complex meso- and
     crystalline
     structures.\cite{Luis-Enrique-Yuri_JPhysCondMat_2014}

     Moreover,
     geometric confinements (e.g., pores and channels) and
     interfacial confinements (e.g., particles  at solid--liquid,
     liquid--liquid, and gas--liquid interfaces) complicate further the
     ordering and phase properties of the
     particles.\cite{Wensink-Lowen..._EurPhysJ_2013,Leferink...Lekkerkerker_EurPhysJ_2013}

     Of
     particular interest are quasi-one-dimensional (q1D) systems, in
     which particles form a necklace-like structure in either
     side-by-side or end-to-end configurations, in such a way that all particles
     are trapped between their first
     neighbors.\cite{Zhang_2006,Liu_2006} In such an environment, and if the interaction range is short enough, the
     phase behavior of the system is very different from that of three-dimensional bulk, since each particle interacts only with its
     first neighbors and the confining
     wall.\cite{Li_2013,Zhang_2014} For example, water molecules
     cannot form H-bonding networks in q1D ultaconfinement, which
     changes several properties of water dramatically, such as the
     conductivity, the diffusivity, and the fluid structure.\cite{Ishai_2020} Similar changes occur in colloidal and
     soft matter systems in q1D confinements, where the shape of the
     particles can be rod-like or
     plate-like.\cite{Xu_2015}

     Therefore, it is not surprising that
     the changes arising in physical and chemical properties make q1D
     systems very attractive for both practical and theoretical
     studies. A practical importance of q1D systems is that
     the necklace-like nanostructures from rod-like and plate-like
     building-blocks of semiconducting nanoparticles have outstanding
     quantum properties, which offer new electric and optical
     applications in  nanotechnology.\cite{Generalova_2021} They
     can also be used in biological and chemical sensing due to their
     anisotropic properties, because they can be embedded into solid
     matrices.\cite{Zhang_2022} The theoretical importance of these
     systems is that some fundamental issues can be addressed by
     studying q1D systems. These issues are, for instance, the existence or
     nonexistence of first-order phase
     transitions,\cite{vanHove_PHYSYCA_1950,Cuesta_JSP_2004}
     glass formation,\cite{Bowles_2000,Semenov_2015}
     jamming,\cite{Kantor_EPL_2009,Bowles_PRL_2009} or the
     possibility of long-range order in low dimensional
     systems.\cite{Hohenberg_1967}

     The minimal model of q1D necklace-like structures is made of hard-body building blocks, where the particles (building blocks) can
     be either spherical or  anisotropic, can rotate to some extent, but
     are restricted to a straight line.\cite{Schwartz_2010} The
     system of strictly confined hard spheres corresponds to a
     one-dimensional (1D) system of hard rods, where the length of the
     rod is equal to the diameter of the sphere. This system belongs
     to the class of exactly solvable models, i.e., the equation of state, the
     pair distribution function, the percolation length, and several
     other thermodynamic and structural properties of 1D hard rods are
     analytical.\cite{Tonks_PhysRev_1936,Salsburg_JCP_1953,Drory_1997,Pugnaloni_1997}
     The general feature of 1D hard rods is that the pressure and the
     pair correlation length diverge at the close-packing density, but
     no phase transition occurs in the entire range of
     density.\cite{Giaquinta_2008}  Interestingly, this system can
     also be realized experimentally to study some dynamical and
     structural properties, such as the diffusion coefficient,
     the structure factor, and the pair correlation
     function.\cite{Bechinger_SCIENCE_2000,Pertsinidis_2005,Diamant_PRL_2005,Bunk_2008,Lin_2009}

     To induce a true phase transition, either {a long-range attractive interaction}
     should be added to the excluded volume
     interactions, like in the case of van der Waals theory,\cite{Kac_1959,Kac_1963} or anisotropic particles
     should be placed on a 1D lattice with some degree of orientational
     freedom.\cite{Casey_1969,Szulga_1987,Saryal_2018,Saryal_2022}
     In general, q1D systems of hard anisotropic particles with
     rotational freedom do not belong to the class of analytically solvable
     models, but the thermodynamic properties and the pair correlation
     functions can be determined exactly by the numerical solution of an
     eigenvalue equation coming from the transfer-matrix
     method.\cite{Lebowitz-Percus-Talbot_JStatPhys_1987,Kantor_2009,Gurin-Varga_PRE_2010,Gurin_PRE_2011}
     In this regard, the exceptions are  q1D additive hard-body
     systems, where the equation of state and the direct correlation
     function can be obtained
     analytically.\cite{Marko_1989,Tejero_1990} Adding some out-of-line positional freedom to the particles can lead to structural
     transitions, jamming, and glassy behavior. In the case of hard
     spheres, a weak fluid-zigzag structural transition takes place
     with increasing
     density\cite{Kofke_JCP_1993,Kamenetskiy-Mon-Percus_JCP_2004,Forster-Mukamel-Posch_PRE_2004,Varga-Ballo-Gurin_JStatMechTheorExp_2011,%
     Gurin-Varga_JCP_2013,Hu_2018,Huerta_2021,Montero_2023,Montero_2023-b}
     and, additionally, the presence of special jammed states shows the existence of
     glass-like structures.\cite{Godfrey-Moore_PhysRevE_2014,Godfrey-Moore_PhysRevE_2015,Robinson-Godfrey-Moore_PhysRevE_2016,Zhang_2020}
     Moreover, the correlation lengths diverge at the close-packing
     density as infinitely long zigzag order
     evolves.\cite{Hu_2021} If the shape of the hard
     particle is anisotropic, the competition between fluid-like and
     solid-like structures gives rise to anomalous structural
     transition, which looks like a first-order phase
     transition.\cite{Gurin_PRE_2016} In addition to this, the phase
     behavior of both spherical and non-spherical hard-body fluids
     becomes very complex {by allowing the particles to pass each other, since} tilted, chiral, and achiral structures
     become the close-packing structure by changing the size of the
     pore.\cite{Fu-et.al_SoftMatter_2017,Jin2020,Jin_2021,Basurto_2021}

     To test the reliability of approximate theoretical methods, such
     as the classical density functional theory (DFT) and the integral-equation approximations, which are devised for two- and
     three-dimensional systems, 1D and q1D systems with short-range
     interactions can be considered as an ideal playground, since
     the output of these theories can be compared with the exact
     results coming from transfer-matrix (TM) and
     neighbor-distribution (ND) methods.  The former approximate
     theories have the advantage that they can be easily generalized
     to higher dimensions, while the extension of exact methods is
     still challenging, even in one dimension, if the pair interaction
     is not restricted to the first neighbor.\cite{Fantoni_2017} The TM and ND methods
     proved to be very successful  for 1D and q1D systems
     with continuous positional and orientational freedom, such as the
     fluid of hard needles\cite{Kantor_EPL_2009,Kantor_2009} and that
     of hard disks.\cite{Kofke_JCP_1993,Montero_2023-b}  Note that
     the freely rotating hard needles can be considered as a simple
     model of liquid crystals in one dimension\cite{Gurin_PRE_2011}
     and even the liquid crystal elastomers can be studied with the
     inclusion of harmonic elastic forces between the neighboring
     needles.\cite{Liarte_2023} Regarding the development of DFT,
     exact functionals have been derived only for hard rods\cite{Percus_1976}
     and hard-rod mixtures,\cite{Vanderlick_1989} while even the
     fundamental-measure density functional is approximate for
     non-additive mixtures.\cite{Schmidt_2007} The problems, failures,
     and challenges in obtaining accurate DFT of non-additive and q1D
     fluids are reviewed in Refs.~\onlinecite{Percus_2002,Mulero_2008}.

     Another
     possibility to study q1D systems is to use the Parsons--Lee (PL)
     theory,\cite{Parsons_1979,Lee_1987} which is also approximate,
     but it proved to be quite accurate for describing the
     orientational ordering properties and the isotropic-nematic
     transition of hard nonspherical particles in two and three
     dimensions.\cite{McGrother_1996,Camp...Kofke_JCP_1996,Varga_2000}
     Interestingly, its success is poorly understood for the equation
     of state and the transition densities of isotropic-nematic phase
     transition. Moreover, to our knowledge, its applicability has not been studied in
     one dimension yet. Therefore, one can get some insight into the
     success of the PL theory by studying some 1D hard-body fluids, where
     the shape of the particle can be both convex and concave.

     In this study, we examine the phase behavior and structural
     properties of q1D hard-body fluids, where the shape of the
     particle is rod-like. The particles are placed into a very narrow
     channel with either rectangular or circular cross section, where
     they form a single-file fluid with only first-neighbor
     interactions. The effect of out-of-line positional freedom is
     completely neglected, i.e., the particles are allowed to move
     freely only in one spatial dimension, while the out-of-line
     orientational freedom is restricted to two and three states in
     rectangular and circular channels, respectively. We show that
     hard prisms behave as an additive mixture, where all components
     have the same chemical potential. Contrary to this, the hard
     dumbbells can be represented as a non-additive mixture with the
     same constraint for the chemical potentials. Both systems exhibit
     orientational ordering with increasing density, where the phase
     is isotropic only at vanishing density. The fluid of hard prisms
     becomes identical with that of hard rods at close packing, where
     the shortest length of the prism along the channel corresponds to
     the diameter of the rod. This system can be characterized by
     diverging positional correlation at close packing, while it
     lacks  orientational correlation. The phase behavior of hard
     dumbbells is very different because it forms crossed structures
     at high densities, where the angle between the neighboring
     particles is 90$^\circ$. Moreover, the dumbbells are more ordered since both the positional and orientational correlation lengths
     diverge at close packing. We show that the PL theory, which is
     devised for isotropic and nematic phases of two- and
     three-dimensional hard-body fluids, is exact for additive q1D
     fluids, regardless of the number of orientations, while it is only approximate for non-additive ones. To
     take into account exactly the effect of non-additive
     interactions, TM or ND methods should be used. In addition, these
     two exact methods complement each other because the ND method
     provides information about the changes occurring in the
     positional order, while the TM method is more suitable to study
     the orientational ordering properties of the system.

The organization of this paper is as follows.
The prism and dumbbell models are presented in Sec.~\ref{sec2}. Then, Sec.~\ref{sec3} is devoted to the  PL theory and  the exact TM and ND methods. The results for the bulk properties, the pair distribution function, and the correlation lengths (both orientational and positional) are presented and discussed in Sec.~\ref{sec4}. Finally, Sec.~\ref{sec5} offers the main conclusions of the paper.

     \section{Models}
\label{sec2}
\begin{figure}
      \includegraphics[width=\columnwidth]{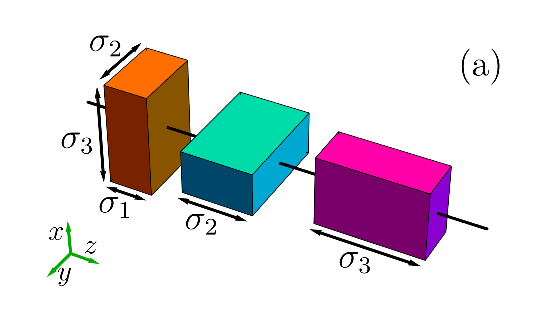}\\
      \includegraphics[width=\columnwidth]{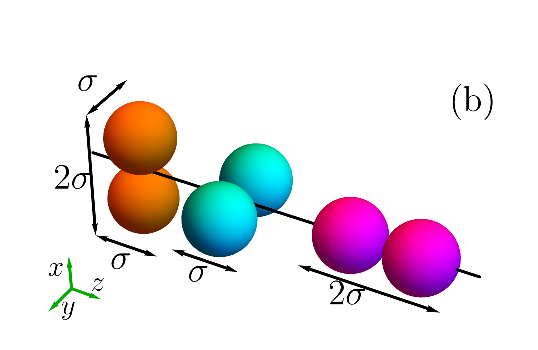}
      \caption{Schematics of (a) hard prisms and (b) hard dumbbells in {q}1D
        channels. The particles are allowed to orient along $x$, $y$,
        and $z$ axis with the corresponding lengths ($\sigma_{1}$, $\sigma_{2}$,
        and $\sigma_{3}$) along the $z$ axis. In the case of
        dumbbells, $\sigma_{1}=\sigma$, $\sigma_{2}=\sigma$ and
        $\sigma_{3}=2 \sigma$.
  \label{fig:model}}
\end{figure}

     We use simple hard-body models for our {q}1D study, where the
     possible orientations of the particles are restricted to two or
     three orientations only, as sketched in Fig.~\ref{fig:model}. We
     assume that the centers of the particles are restricted to the
     $z$ axis, but the particles can move freely along this axis. The
     particles are not allowed to overlap as they are hard objects. We
     can see in Fig.~\ref{fig:model} that the occupied length of the particle can be
     $\sigma_{1}$, $\sigma_{2}$, and $\sigma_{3}$ along the $z$ axis as the
     particle can orient its largest length along the $x$, $y$, and $z$ axes, respectively. For the sake of
     simplicity, we assume that $\sigma_{1} \leq \sigma_{2} \leq
     \sigma_{3}$ for hard prisms and $\sigma_{1}=\sigma,
     \sigma_{2}=\sigma$, and $\sigma_{3}=2 \sigma$ for hard
     dumbbells. We measure all lengths and make all quantities
     dimensionless with $\sigma_1$, which is then the unit of length. The
     hard prisms are additive, because the contact distance between
     two prisms is given by $\sigma_{i j} =
     \left(\sigma_{j}+\sigma_{j}\right) / 2$ for any pair of
     orientations $(i, j=1,2,3)$. This is not true for hard dumbbells
     because $\sigma_{i j} \neq\left(\sigma_{j}+\sigma_{j}\right) / 2$
     for $i \neq j$; more specifically, $\sigma_{12}/\sigma=1/\sqrt{2}\simeq 0.707$ and $\sigma_{13}/\sigma=\sigma_{23}/\sigma=(1+\sqrt{3})/2\simeq 1.366$. With these two models we can examine the effects
     of additive and  non-additive hard-body interactions.

     \section{Theory}
\label{sec3}

     In this section, we present three different theories for q1D hard-body fluids. In the three cases, the number $n$ of internal states (or, in the mixture language, the number of components) is arbitrary. So are the lengths $\sigma_i$, the cross contact distances $\sigma_{ij}$, and the nature (additive or non-additive) of the interactions.

     We start with the PL theory, which provides the
     equation of state and the fraction of particles
     $\left(x_{i}\right)$ having length $\sigma_{i}$ along the $z$
     axis. Then, we present the TM method for the exact calculation of
     the Gibbs free energy and other thermodynamic properties. In
     addition, we show that this method provides information about the
     orientational correlations along the $z$ axis. Finally, we derive the
     exact equations for the orientation-dependent pair distribution
     function using the ND method.

     \subsection{Parsons--Lee (PL) theory}
\label{sec3.1}

     The easiest way to derive the PL theory from the
     three possible ways\cite{Vroege_1992,Groh_1997,Padilla_1997} is
     to start from the virial series of the excess free energy
     density $F_\ex$,
\begin{equation}
     \frac{\beta F_{\ex}}{L}
  =  \sum_{i=2}^{\infty} \frac{ B_{i}}{i-1} \rho^{i},
 \label{eq:virial}
\end{equation}
     where $\beta=1 /k_{B} T$ is the inverse temperature ($k_B$ being the Boltzmann constant),
     $L$ is the length of the channel, $B_{i}$ is the $i$th virial
     coefficient, $\rho=N / L$ is the linear number density, and $N$
     is the number of particles in the channel. According to  Lee's idea,\cite{Lee_1987} a mapping procedure can be made between
     the system of anisotropic particles and that of spherical
     particles through the virial coefficients. In the case of $1
     \mathrm{D}$ confinement, the hard rods having length $d$ can be
     used in the mapping procedure as follows,
\begin{equation}
     B_{i}
  \approx \frac{B_{2}}{B_{2}^{\RR}} B_{i}^{\RR},
 \label{eq:virial_Lee}
\end{equation}
     where $B_{i}^{\RR}$ denotes the $i$th virial coefficient of the hard
     rods. This equation assumes that the virial coefficients of the
     hard rods and those of anisotropic particles are proportional to
     each other. {Substituting Eq.~(\ref{eq:virial_Lee}) into Eq.~(\ref{eq:virial}), we obtain, after simplification, that}
\begin{equation}
     F_{\ex}
  =   F_{\ex}^{\RR} \frac{B_{2}}{B_{2}^{\RR}},
 \label{eq:F_Lee}
\end{equation}
     where $F_{\ex}^{\RR}$ is the excess free energy
      of the hard rods. Due to Tonks,\cite{Tonks_PhysRev_1936}
     this free energy term is analytically known as
\begin{equation}
     \frac{\beta F_{\ex}^{\RR}}{L}
  = -\rho \ln (1-\rho d).
 \label{eq:F_exc_rod}
\end{equation}

     Using an $n$-state representation of the possible orientations of
     the hard-body anisotropic particles, it can be shown that the second virial
     coefficient is given by
\begin{equation}
     B_{2}
  =   \sum_{i, j=1}^{n} x_{i} x_{j} \sigma_{i j}.
 \label{eq:B_2}
\end{equation}
     In the case of hard rods,
     Eq.~(\ref{eq:B_2}) simplifies to $B_{2}^{\RR}=d$. Thus, Eqs.~\eqref{eq:F_Lee} and \eqref{eq:F_exc_rod} yield
\begin{equation}
     \frac{\beta F_{\ex}}{L}
  = -\rho \ln (1-\rho d)\; \frac{1}{d} \sum_{i, j=1}^{n} x_{i} x_{j} \sigma_{i j}.
 \label{eq:F_exc_rod_Lee}
\end{equation}
     To complete the determination of the excess free energy, a relationship
     between the hard-rod length $d$ and the lengths $\{\sigma_{ij}\}$ of the anisotropic
     particles is needed. A natural choice is  $d=\langle\sigma\rangle=\sum_{i=1}^{n}
     x_{i} \sigma_{i}$, which ensures that the occupied length of the
     rods and that of the anisotropic particles are the same along the $z$
     axis in the case of additive cross interactions. Here we note that the generalization of
     Eq.~(\ref{eq:F_exc_rod_Lee}) is straightforward in higher
     dimensions as the hard disk and sphere can be used as a reference
     mapping particle in two and three dimensions, respectively. The
     total free energy, which is the sum of ideal and excess
     terms ($F=F_{\id}+F_{\ex}$),
     can be written as
\begin{equation}
     \frac{\beta F}{L}
  =  \sum_{i=1}^{n} \rho_{i} \left(\ln \rho_{i} - 1\right)
    -\frac{\ln (1-\eta)}{ \eta} \sum_{i, j=1}^{n} \rho_{i} \rho_{j} \sigma_{i j},
 \label{eq:F_rod_Lee}
\end{equation}
     where  $\rho_{i}=\rho x_{i}$ is the  density of component $i$ and
     $\eta=\sum_{i=1}^{n} \rho_{i} \sigma_{i}=\rho d=\rho\langle\sigma\rangle$ is the linear packing
     fraction. Here, without loss of generality, we have taken the thermal de Broglie wavelength equal to unity. The chemical potential of component $i$ and the
     pressure can be obtained from Eq.~(\ref{eq:F_rod_Lee}) using
     standard thermodynamic relations as follows,
     \begin{subequations}
     \label{eq:mu&P}
\begin{equation}
     \beta \mu_{i}
  =  \frac{\partial\left(\beta F/L\right)}{\partial \rho_{i}}, \quad(i=1, \ldots n),
 \label{eq:mu}
\end{equation}
\begin{equation}
     \beta P
  = -\frac{\beta F}{L}+\sum_{i=1}^{n} \beta \mu_{i} \rho_{i}.
 \label{eq:P}
\end{equation}
     \end{subequations}

     In the case of additive excluded length, one has  $\sum_{i, j=1}^{n} \rho_{i} \rho_{j} \sigma_{i j}=\rho\eta$, so that the total free energy simplifies to
\begin{equation}
     \frac{\beta F}{L}
  =  \sum_{i=1}^{n} \rho_{i}\left( \ln \rho_{i}-1\right)-\rho \ln (1-\eta) .
 \label{eq:F_additive}
\end{equation}
     From this equation one gets the chemical potentials and the
     pressure using Eqs.~(\ref{eq:mu&P}), i.e.,
     \begin{subequations}
     \label{eq:mu&P_additive}
\begin{equation}
     \beta \mu_{i}
  =  \ln \rho_{i} -\ln (1-\eta)+\frac{\rho \sigma_{i}}{1-\eta},
 \label{eq:mu_additive}
\end{equation}
\begin{equation}
     \beta P
  =  \frac{\rho}{1-\eta}.
 \label{eq:P_additive}
\end{equation}
  \end{subequations}
    The fraction of particles can be obtained from the condition that
     the chemical potential of all components are fixed to a given
     value $(\mu)$,\cite{Onsager_AnnNYAcadSci_1949} i.e., $\mu=\mu_{1}=\cdots=\mu_{n}$. Therefore, in the case of additive interactions, we have
\begin{equation}
     x_{i}
  =  \frac{e^{-\rho \sigma_{i} /(1-\eta)}}{\sum_{j=1}^{n} e^{-\rho \sigma_{j} /(1-\eta)}}.
 \label{eq:x(rho)_additive}
\end{equation}
     Note that this  actually represents a set of transcendental equations for $\{x_i\}$, since $\eta$ depends on {$\{x_i\}$}. Thus, it is not possible to provide a closed formula for $x_{i}$ as
     a function of $\rho$. However, the density dependence
     of Eq.~(\ref{eq:x(rho)_additive}) can be replaced with the
     pressure one using Eq.~(\ref{eq:P_additive}). The result is
\begin{equation}
     x_{i}
  =  \frac{e^{-\beta P \sigma_{i}}}{\sum_{j=1}^{n} e^{-\beta P \sigma_{j}}}.
 \label{eq:x(P)_additive}
\end{equation}
     Using Eqs.~(\ref{eq:mu&P_additive}), one
     can express the chemical potential, density, and packing fraction
     as functions of pressure:
     \begin{subequations}
     \label{eq:mu&rho&eta(P)_additive}
\begin{align}
     \beta \mu
 &=  \ln (\beta P )-\ln
     \sum_{i=1}^{n} e^{-\beta P \sigma_{i}},
 \label{eq:mu(P)_additive}
 \\
     \rho^{-1}
 &=  \frac{\sum_{i=1}^{n} \sigma_{i} e^{-\beta P \sigma_{i}}}
          {\sum_{i=1}^{n} e^{-\beta P \sigma_{i}}}+\frac{1}{\beta P},
 \label{eq:rho(P)_additive}
\end{align}
\begin{equation}
     \eta
  =  \frac{\beta P \sum_{i=1}^{n} \sigma_{i} e^{-\beta P \sigma_{i}}}
          {\sum_{i=1}^{n}\left(1+\beta P \sigma_{i}\right) e^{-\beta P \sigma_{i}}}.
 \label{eq:eta(P)_additive}
\end{equation}
     \end{subequations}
     These results show that the pressure is the natural input of 1D
     additive systems, which is consistent with the TM and ND methods.

      It is worth mentioning  that Eqs.~\eqref{eq:mu&rho&eta(P)_additive} can be
     reproduced from the virial theorem using the decoupling
     approximation of the positional and orientational degrees of
     freedom.\cite{Vroege_1992} In this approximation, the following
     three steps are used: (1) the pair potential of anisotropic hard
     bodies $\left(u_{i j}\right)$ is scaled into that of hard rods,
     i.e., $u_{i j}(z) = u\left(z d/\sigma_{i j}\right)$, where $z$ is
     the distance between two particles, (2) the pair distribution
     function can be determined from $g_{i j}(z, \rho) \approx
     g^{\RR}\left(z d/ \sigma_{i j}, \rho\right)$, where $g^{\RR}$ is the
     pair distribution function of hard rods, and (3) the occupied distance of the particles
     equals to that of hard rods.

     We finally
     note that the PL theory for the non-additive case,
     i.e., $\sigma_{i j} \neq (\sigma_{i}+\sigma_{j})/2$, does not produce
     analytical results for  the fractions $x_{i}$ and the thermodynamic
     quantities.

     \subsection{Transfer-matrix (TM) method}
\label{sec3.2}

     The simplest way to determine the partition function and the
     derived thermodynamic quantities of {q}1D systems is to work in the
     isothermal-isobaric ensemble, where the partition function can be
     factorized as a { product of matrices given by}
\begin{equation}
     K_{i j}
  =  \frac{e^{-\beta P \sigma_{i j}}}{\beta P},
 \label{eq:K}
\end{equation}
     where, as said before, $\sigma_{i j}$ is the contact distance between two
     neighboring particles having $i$ and $j$ orientations,
     respectively.  According to the TM method, the important quantities are the eigenvalues and the
     corresponding eigenvectors of the matrix $K_{ij}$. In the $n$-state
     system, the eigenvalue equation is given by
\begin{equation}
     \sum_{j=1}^{n} K_{i j} \psi_{j}^{(k)}
  =  \lambda_{k} \psi_{i}^{(k)},
 \label{eq:eigenvalue_eq}
\end{equation}
     where $\lambda_{k}$ is the $k$th eigenvalue, while
     $\psi_{i}^{(k)}$ is the $i$th component of the corresponding
     eigenvector. One gets the eigenvalues from the condition that the
     determinant of the matrix {$K_{ij}-\lambda\delta_{ij}$} must be zero, while the
     corresponding eigenvectors are obtained from the eigenvalue equation,
     Eq.~(\ref{eq:eigenvalue_eq}).
     If the eigenvectors are normalized, then
     \beq
     \label{eq:lambda_k}
     \lambda_k=\sum_{i,j=1}^n K_{ij}\psi_{i}^{(k)}\psi_{j}^{(k)}.
     \eeq
     From here, it is easy to prove\cite{Montero_2023} that
     \beq
     \label{eq:Deriv_lambda}
     \frac{\partial (\beta P\lambda)}{\partial \beta P}=-\beta P\sum_{i,j=1}^n K_{ij}\sigma_{ij}\psi_{i}\psi_{j} ,
     \eeq
     {where $\lambda=\max \left(\lambda_{1}, \ldots,
       \lambda_{n}\right)$ and $\psi_i$ is the corresponding
       eigenvector.}

     The Gibbs free energy can be
     obtained from $\beta G / N=-\ln \lambda$. The equation of
     state, which connects the pressure and the density, is given by
\bal
     \frac{1}{\rho}
  =&  \frac{\partial \beta G / N}{\partial \beta P}
  = -\frac{1}{\lambda} \frac{\partial \lambda}{\partial \beta P}\nn
  =&\frac{1}{\beta P}+\frac{1}{\lambda}\sum_{i,j=1}^n K_{ij}\sigma_{ij}\psi_{i}\psi_{j},
 \label{eq:rho_TMM}
\eal
where {Eq.~\eqref{eq:Deriv_lambda} has been used}.
     Further information about the ordering can be gained from
     the (normalized) eigenfunction $\psi_{i}$, since the fraction of particles having a  length $\sigma_{i}$
     along the $z$ axis is  $x_i=\psi_{i}^2$.  Moreover, the orientational
     correlation between two pairs can be characterized with the help
     of the orientational correlation length ($\xi_{o}$),\cite{Gurin_PRE_2011} which
     is obtained from the two largest eigenvalues as
\begin{equation}
     \xi_{o}^{-1}
  =  \ln \frac{\lambda}{\left|\lambda^{*}\right|},
 \label{eq:xi}
\end{equation}
     where $\lambda^{*}$ is the second largest eigenvalue (in absolute value) of $K_{ij}$.

     In the additive case, i.e., $\sigma_{i
       j}=\left(\sigma_{i}+\sigma_{j}\right)/2$, the eigenvalues and
     the eigenvectors can be obtained easily, because the matrix
     elements can be factorized as follows,
\begin{equation}
     K_{i j}
  =  \sqrt{K_{i}K_{j}},
 \label{eq:K_additive}
\end{equation}
     where $K_{i}=K_{ii}=e^{-\beta {P} \sigma_{i}}/\beta P$. Inserting
     Eq.~(\ref{eq:K_additive}) into Eq.~(\ref{eq:eigenvalue_eq}) one
     gets that
\begin{equation}
    \lambda_k \psi_{i}^{(k)}
  =  \sqrt{K_{i}} \sum_{j=1}^{n} \sqrt{K_{j}} \psi_{j}^{(k)}.
 \label{eq:psi_additive}
\end{equation}
          {Actually, Eq.~(\ref{eq:K_additive}) expresses that the matrix
       $K_{ij}$ is the Kronecker product of a vector $\sqrt{K_{i}}$ by
       itself. As a consequence, apart from a constant multiplier,
       $K_{ij}$ is the matrix of a rank $1$ projector, and therefore, all its
       eigenvalues are zero except one, which is the largest,
       $\lambda=\sum_{i=1}^{n} K_{i}$; moreover, the corresponding
       eigenvector is proportional to the vector $\sqrt{K_{i}}$. From}
     the normalization condition we have
\begin{equation}
     x_{i}
  =  \frac{K_{i}}{\sum_{j=1}^{n} K_{j}}
  =  \frac{e^{-\beta P \sigma_{i}}}{\sum_{j=1}^{n} e^{-\beta P \sigma_{j}}},
 \label{eq:x_i_TMM}
\end{equation}
     which is identical to Eq.~(\ref{eq:x(P)_additive}) of the PL
     theory. It is easy to show that the TM method provides the same
     results for the equation of state, the packing fraction, and the
     chemical potential as that of the PL theory for arbitrary $n$ and $\{\sigma_i\}$  provided the interactions are additive.  As the TM method
     is exact, it turns out that the PL theory is also an exact theory
     for {q}1D $n$-state hard-body systems in the additive case. At this
     point it is worth noting that the PL theory provides only the
     thermodynamic properties, while the TM method can be used to
     determine the structural properties, too. For example,
     Eq.~(\ref{eq:xi}) shows that there is no orientational
     correlation between the particles, i.e., the orientational correlation
     length is zero for additive systems because $\lambda^*=0$.

     As
     far as  non-additive systems are concerned, the matrix element $K_{i
       j}$ cannot be factorized as a product of two one-body terms, as
     in Eq.~(\ref{eq:K_additive}). Therefore, the resulting equations are
     more complicated for the ordering and thermodynamic properties,
     since the determinant of the matrix $K_{i j}-\lambda\delta_{ij}$ becomes an $n$th-order polynomial with nonzero eigenvalues $\lambda_{1}, \ldots, \lambda_{n}$. Using these exact results, we will show that the PL
     theory is not exact for non-additive systems.

     \subsection{Neighbor-distribution (ND) method}

The complete determination of the physical properties of the system requires the knowledge of the pair distribution function $g_{ij}(z)$, which is proportional to the probability of finding a particle at position $z$ and with orientation $j$, given that a particle with orientation $i$ is located at the origin ($z=0$).

     To calculate
     $g_{ij}(z)$ in q1D systems, one needs to use the isothermal-isobaric ensemble and start from the determination of the nearest-neighbor probability
     distribution function, from which the $\ell$th-neighbor distribution function can be obtained by iterated convolutions. For this reason, here we will refer to this methodology as the ND method. In addition to $g_{ij}(z)$, the equation of state and other
     thermodynamic quantities can be calculated with it, yielding
     exactly the same results as those from the {TM} method. A
     drawback of the ND method is, however,  that it does not
     provide information about the orientational correlations.

     In this section, we summarize the main results derived from the ND method and refer the reader to Chap.~5 of Ref.~\onlinecite{Santos_2016} and, especially, Sec.~III of Ref.~\onlinecite{Montero_2023-b} for further details.
The exact pair distribution function is
\beq
\label{eq:gx_ij}
g_{ij}(z)=\frac{1}{\rho\sqrt{x_ix_j}}\sum_{\ell=1}^{\lfloor z/\sigma_{\min} \rfloor} \frac{Q_{ij}^{(\ell)}(z)}{\lambda^\ell},
\eeq
where
\begin{subequations}
\label{eq:qijn_expl}
\beq
Q_{ij}^{(1)}(z)=R^{(1)}(z;\sigma_{ij}),
\eeq
\beq
Q_{ij}^{(\ell)}(z)=\sum_{k_1=1}^n\sum_{k_2=1}^n\cdots\sum_{k_{\ell-1}=1}^nR^{(\ell)}(z;\Sigma_{i k_1k_2\cdots k_{\ell-1}j}),\quad \ell\geq 2,
\eeq
\end{subequations}
with
\begin{subequations}
\beq
\label{3.25a}
\Sigma_{i k_1k_2\cdots k_{\ell-1}j}\equiv \sigma_{i k_1}+\sigma_{k_1k_2}+\cdots+\sigma_{k_{\ell-1}j},
\eeq
\beq
\label{eq:qikj}
R^{(\ell)}(z;\alpha)\equiv\frac{e^{-\beta P z}}{(\ell-1)!}(z-\alpha)^{\ell-1}\Theta(z-\alpha).
\eeq
\end{subequations}
In the upper summation limit of Eq.~\eqref{eq:gx_ij}, $\lfloor \cdots \rfloor$ denotes the floor function   and $\sigma_{\min}=\min\{\sigma_{ij}\}$.
Note that $g_{ij}(z)$ presents a jump at $z=\sigma_{ij}$, kinks at $z=\Sigma_{ik_1j}$ ($k_1=1,\ldots,n$), and, in general, singularities of order $\ell-1$ at $z=\Sigma_{i k_1k_2\cdots k_{\ell-1}j}$.

The Laplace transform $\widetilde{G}_{ij}(s)=\int_0^\infty dz\, e^{-sz} g_{ij}(z)$ is given by\cite{Montero_2023-b}
\beq
\label{3.5cc}
\widetilde{G}_{ij}(s)=\frac{1}{\lambda \rho\sqrt{x_ix_j}}\left({\mathsf{\Omega}}(s+\beta P)\cdot\left[\mathsf{I}-\lambda^{-1}{\mathsf{\Omega}}(s+\beta P)\right]^{-1}\right)_{ij},
\eeq
where $\mathsf{\Omega}(s)$ is the $n\times n$ matrix with elements $\Omega_{ij}(s)=e^{-s\sigma_{ij}}/s$. Note that $K_{ij}=\Omega_{ij}(\beta P)$.

In the additive case, $\Omega_{ij}(s)=\sqrt{\Omega_i(s)\Omega_j(s)}$, where $\Omega_i(s)=\Omega_{ii}(s)$, and this simplifies  Eq.~\eqref{3.5cc}. After simple algebra, one finds
\bal
\label{eq:G_additive}
\widetilde{G}_{ij}(s)=&\frac{1}{\rho\sqrt{x_ix_j}}\frac{\Omega_{ij}(s+\beta P)}{\lambda-\sum_{k=1}^n \Omega_k(s+\beta P)}\nn
=&\frac{1}{\rho}\frac{\beta P}{s+\beta P}\frac{e^{-s\sigma_{ij}}}{1-\lambda^{-1}\sum_{k=1}^n \Omega_k(s+\beta P)},
\eal
where in the second step we have taken into account that $x_i=e^{-\beta P\sigma_i}/\lambda \beta P$ in the additive case. The second equality in Eq.~\eqref{eq:G_additive} implies that all the pair distribution functions $g_{ij}(z)$ for additive systems are the same if the origin is shifted to $z=\sigma_{ij}$,\cite{Grodon_2004,Grodon_2005,Schmidt_2007} i.e.,
\beq
\label{eq:shift}
g_{ij}(z)=f(z-\sigma_{ij}),
\eeq
where the function $f(z)$ is common for all pairs.
Also in the additive case, Eqs.~\eqref{eq:gx_ij} and \eqref{eq:qijn_expl} can still be used, but Eq.~\eqref{3.25a} simplifies to
\beq
\label{3.29}
\Sigma_{i k_1k_2\cdots k_{\ell-1}j}=\sigma_{ij}+ \sigma_{k_1}+\sigma_{k_2}+\cdots+\sigma_{k_{\ell-1}}.
\eeq

The asymptotic decay of $g_{ij}(z)-1$ is characterized by the nonzero poles of $\widetilde{G}_{ij}(s)$, i.e., the roots (different from $s=0$) of the determinant of the matrix $\mathsf{I}-\lambda^{-1}{\mathsf{\Omega}}(s+\beta P)$ for non-additive systems [see Eq.~\eqref{3.5cc}] or of $1-\lambda^{-1}\sum_{k=1}^n \Omega_k(s+\beta P)$ for additive systems [see Eq.~\eqref{eq:G_additive}].
If we denote by  $s_\pm=-\kappa\pm\imath\omega$ the pair of conjugate poles  with the  real part closest to the origin, its residue being $|\mathcal{A}_{ij}|e^{\pm\imath\delta_{ij}}$, then, for asymptotically large $z$,
\beq
\label{hij}
g_{ij}(z)-1\approx 2|\mathcal{A}_{ij}|e^{-\kappa z}\cos(\omega z+\delta_{ij}).
\eeq
Thus,  $\xi_p=\kappa^{-1}$  represents the positional {correlation length}, whereas $\omega$ is the (angular) oscillation frequency. As density increases, the imaginary part ($\omega$) can experience a discontinuous jump at a certain density, giving rise to a {structural crossover} from oscillations with a certain frequency  to oscillations with a different one. The origin of this jump resides in the crossing of the real part of two competing poles with different imaginary parts.

     \section{Results}\label{sec4}

     In this section, we present our results for the phase behavior and
     structural properties of hard prisms and dumbbells in {q}1D channels (see Fig.~\ref{fig:model})
     using the PL theory, as well as the TM and ND methods. The particles are allowed
     to move freely along the channel, but their orientational freedom
     is restricted to either two $(n=2)$ or three $(n=3)$ orientations. The main difference between hard prisms and dumbbells is
     that the contact distance between two prisms is additive for any
     pairs of orientations, i.e., $\sigma_{i
       j}=\left(\sigma_{j}+\sigma_{j}\right) / 2$, while hard
     dumbbells are non-additive as $\sigma_{i j}
     \neq\left(\sigma_{j}+\sigma_{j}\right) / 2$ for $i \neq j$
     orientations. We use the following dimensionless quantities:
     $z^{*}=z / \sigma_{1}$, $\rho^{*}=\rho \sigma_{1}$, and
     $P^{*}=\beta P \sigma_{1}$.

     \subsection{Bulk properties}

     \subsubsection{Hard prisms}

\begin{figure}
      \includegraphics[width=\columnwidth]{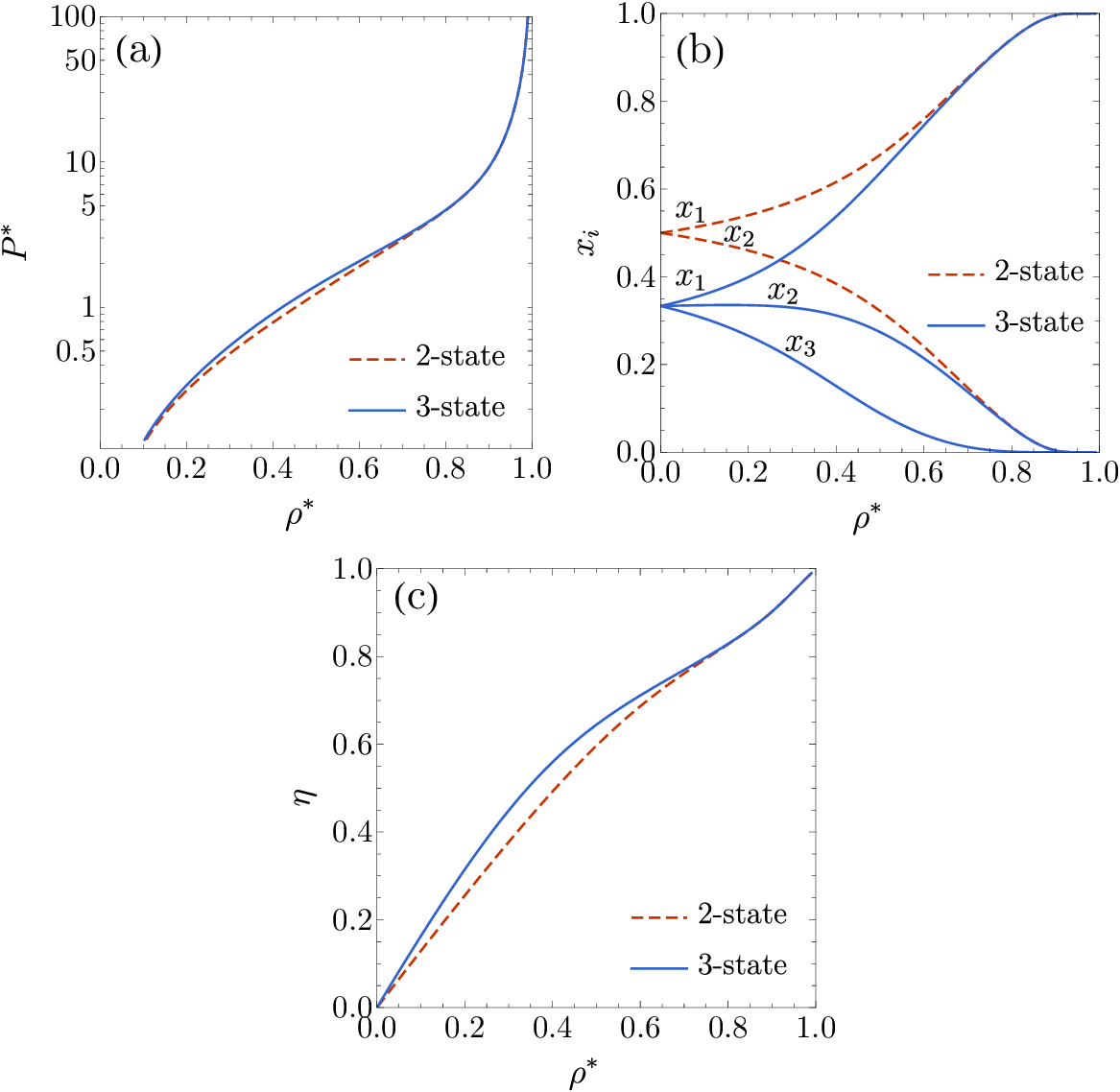}
      \caption{Phase behavior of hard prisms in a {q}1D channel:
        (a) pressure,  (b) mole fraction, and (c) packing fraction as
        functions of density. Particles can orient along the $x$ and
        $y$ axes only in the two-state model, while the $x$, $y$, and $z$ axes
        are allowed in the three-state one. The lengths of the prism are
        chosen as follows: $\sigma_{1}=1$, $\sigma_{2}=1.6$,  and
        $\sigma_{3}=2.4$. The pressure and density are dimensionless:
        $P^{*}=\beta P \sigma_{1}$ and $\rho^{*}=\rho \sigma_{1}$.
  \label{EOS_etc_prism}}
\end{figure}

     We start with the simple {q}1D fluid of hard prisms and compare the
     bulk properties of two-state and three-state models. It is easy to show
     that the prisms are parallel with their shortest length
     $\sigma_{1}$ along the $z$ axis at close packing. Therefore, they
     must behave as a 1D fluid of hard rods at high densities, where
     $x_{1}\to 1$ and $\eta\to \rho \sigma_{1}$
     describe the phase properties of the system. In the low-density, ideal-gas
     limit, the particles form an isotropic phase in both models, with
     $x_{i}\to 1 / n$ holding for the $n$-state model. This can be
     obtained from Eq.~(\ref{eq:x(rho)_additive}) or
     Eq.~(\ref{eq:x_i_TMM}) by taking the limits $\rho \rightarrow 0$ or $\beta P
     \rightarrow 0$, respectively. We can see from these limiting results
     that the structure of hard prisms changes from isotropic to a
     perfect nematic fluid with increasing density. The results of
     Eqs.~(\ref{eq:x(rho)_additive}), (\ref{eq:rho(P)_additive}) and
     (\ref{eq:eta(P)_additive}) are shown together in
     Fig.~\ref{EOS_etc_prism}.
     Starting with the equation of state ($P^*$ vs $\rho^*$), we
     can see in Fig.~\ref{EOS_etc_prism}(a) that there are some differences in the resulting curves
     at intermediate densities, while the two-state and three-state prisms
     behave almost identically at very low and high densities. The low-density agreement is trivial because of the ideal-gas limit, but
     the high-density one is due to the orientational ordering of the
     prisms into the state with  length $\sigma_{1}$ along the $z$ axis.
     This can be seen clearly in Fig.~\ref{EOS_etc_prism}(b),
     where $x_{1}$ goes to 1 with increasing density for both 2- and
     three-state models. The effect of increasing density is that
     neighboring  particles get closer to each other, which reduces
     the available room for the particles. To minimize the
     translational entropy loss, the particles reduce their length
     along the $z$ axis with orientational ordering, which manifests
     in an orientational entropy loss. Therefore, the competition
     between the translational and orientational entropies results in
     a continuous structural change from the isotropic to the
     perfectly ordered nematic fluid. It can be seen in
     Fig.~\ref{EOS_etc_prism}(b) that the ordering is more pronounced
     for the three-state model than for the two-state one, since the
     translational entropy gain is higher
     as the number of particles having  length $\sigma_{3}$ (which is the longest
     side of the prism) decreases. As  the orientation with length $\sigma_{3}$ is missing in the
     two-state model, the changes are smoother in the two-state model than in the
     three-state one.

     It is also obvious that the pressure and the packing
     fraction are higher in the three-state model than in the two-state one
     at a given density [see Figs.~\ref{EOS_etc_prism}(a) and \ref{EOS_etc_prism}(c)]
     because the particles are always closer to each other in the
     three-state model since the orientational entropy makes $x_{3}\neq 0$. The
     difference between the two models virtually disappears at
     $\rho^*=0.8$, because $x_{3}$ is almost zero beyond this
     density. Moreover, both systems become almost a 1D fluid of hard
     rods with length $\sigma_{1}$ for $\rho^*>0.9$, where $x_{2}$ and
     $x_{3}$ are practically zero. Note that the equation of state of
     $1 \mathrm{D}$ hard rods of length $d$, which
     is given by $\beta P=\rho /(1-\eta)$ with
     $\eta=\rho d$,\cite{Tonks_PhysRev_1936} can
     describe  $n$-state additive hard-body systems, such as the 2- and
     three-state prisms, if $d=\langle\sigma\rangle$ is the average length of
     the particle along the $z$ axis, as done in the PL theory [see Eq.~\eqref{eq:P_additive}]. In the light of this result, the curve $\eta$
     vs $\rho^*$  measures the deviation from 1D hard rods,
     because the curve is a straight line for $1
     \mathrm{D}$ hard rods since $\eta=\rho
     \sigma_{1}$ in that case. Figure~\ref{EOS_etc_prism}(c) shows that the deviation
     from the $1 \mathrm{D}$ hard-rod system increases with the number of
     orientational states mainly at intermediate densities, while the
     $n$-state system converges toward the 1D hard-rod system at very high
     densities, where $\langle\sigma\rangle \approx \sigma_{1}$. These
     results show that the addition of more and more out-of-line
     orientational freedom to the system increases the deviation from
     the 1D fluid of hard rods.

     \subsubsection{Hard dumbbells}

     The phase behavior of {q}1D hard dumbbells is more complicated due
     to the presence of non-additive interactions. As the two hard spheres making a dumbbell
     are in contact  (see Fig.~\ref{fig:model}),
     the length of the particle can be either $\sigma$ (in states 1 and
     2) or $2 \sigma$ (in state 3) along the $z$ axis. Keeping in mind
     the symmetry property of the contact distance
     ($\sigma_{ij}=\sigma_{ji}$), one can get all $\sigma_{ij}$ using
     the following special values: $\sigma_{11}=\sigma_{22}=\sigma$,
     $\sigma_{33}=2\sigma$, $\sigma_{12}=\sigma/\sqrt{2}$, and
     $\sigma_{13}=\sigma_{23}=(1+\sqrt{3})\sigma/2$. These contact
     distances and the pressure are the inputs of the exact TM and ND
     methods. Note that the input of the PL theory is the density and the
     effective length $d=\langle\sigma\rangle$. We examine the
     following 2- and three-state systems: (a) a two-state model in which the state 1 with length $\sigma$
     and the state 3 with length $2 \sigma$  are allowed,  and (b)  a three-state model where all states are included.

     To understand the phase behavior of the q1D dumbbell fluid,
     it is worth considering the close-packing structure of the
     system. The possible shortest distance between two dumbbells is
     $\sigma$ in the two-state model, i.e., one dumbbell occupies a distance
     $\sigma$  and the close-packing density is given by
     $\rho_{\text{cp}}=1 / \sigma$. Therefore, the dumbbells are parallel at
     close packing and form a perfect nematic order with
     $x_{1}=1$. Consequently, the dumbbells behave as a $1 \mathrm{D}$
     hard-rod fluid at high densities with $\beta P=\rho /(1-\eta)$ and
     $\eta=\rho \sigma$. This shows that the close-packing behavior
     of two-state dumbbells and that of prisms are the same. In the
     three-state model, however, the shortest distance between two dumbbells is
     given by $\sigma / \sqrt{2}$, i.e., the neighboring dumbbells
     must be perpendicular with respect to each other and both perpendicular to the $z$ axis at close
     packing. This means that one dumbbell occupies a distance $\sigma /
     \sqrt{2}$ and the close-packing density is given by
     $\rho_{\text{cp}}=\sqrt{2} / \sigma$. This ordered structure is not
     planar nematic, because the order of particles with states 1 and
     2 is not random along the $z$ axis. Note that if two particles are
     parallel, the shortest distance between them is $\sigma$, which
     is higher than $\sigma / \sqrt{2}$. Therefore, the close-packed
     structure of the three-state dumbbell system is the sequence
     1-2-1-2-$\cdots$ of the states along the $z$ axis. We use
     the name ``crossed'' for this ordered structure because neighboring
     particles like to be perpendicular to each other. The consequence
     of the crossed ordering for the close-packing properties is that
     $x_{1}=x_{2}=1 / 2$ and $\eta_{\text{cp}}=\sqrt{2}$.

     From these results, we can see that, whereas
     $\eta=\rho \sum_{i=1}^{n} x_{i} \sigma_{i}$ is the real 1D
     packing fraction of additive systems,  it is just a density-dependent quantity for non-additive ones. This fact has a serious
     consequence for the applicability of the PL theory because the
     PL excess free energy diverges at $\eta=1$ [see
     Eq.~(\ref{eq:F_rod_Lee})], which is below the maximal value
     ($\eta_{\text{cp}}=\sqrt{2}$) for the three-state dumbbell model. Therefore, the PL theory is
     not exact and is unable to predict the phase behavior of hard
     dumbbells if  $\eta>1$.

\begin{figure}
      \includegraphics[width=\columnwidth]{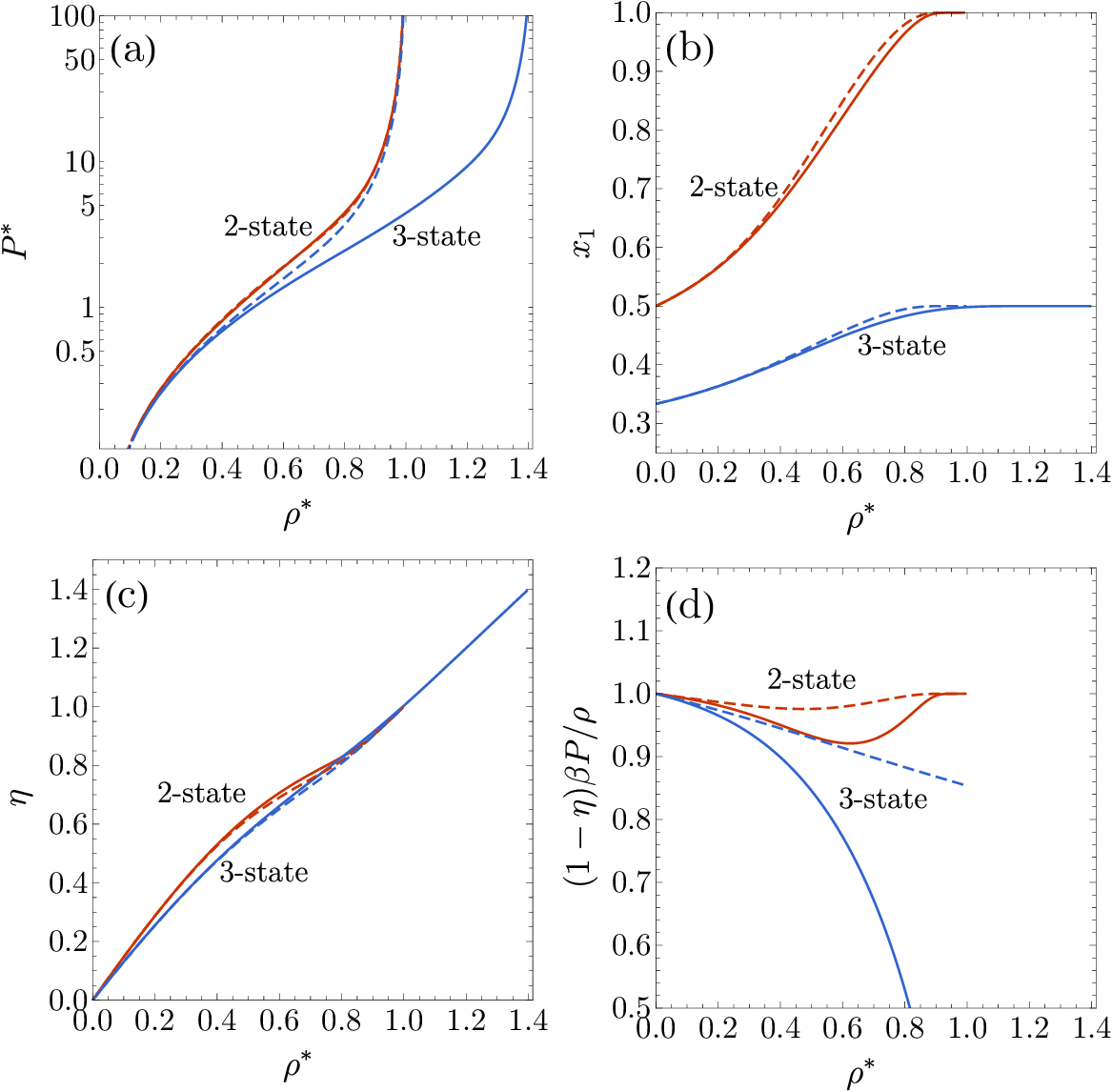}
      \caption{Phase behavior of hard dumbbells in a {q}1D
        channel: (a) pressure, (b) mole fraction, (c) packing fraction, and (d) $(1-\eta)\beta P / \rho$ as
        functions of the reduced density. Particles can orient along the $x$
        and $z$ axes in the two-state model, while the $x$, $y$, and $z$ axes are
        allowed in the three-state one. The dashed curves correspond to
        the results of the PL theory, while the solid curves are
        the exact results. The lengths of the dumbbells are chosen as
        follows: $\sigma_{1}=\sigma$, $\sigma_{2}=\sigma$, and
        $\sigma_{3}=2 \sigma$. The pressure and density are
        dimensionless: $P^{*}=\beta P \sigma_{1}$ and $\rho^{*}=\rho
        \sigma_{1}$. The corresponding close-packing densities are
        given by $\rho_{\text{cp}}^{*}=1$ and $\rho_{\text{cp}}^{*}=\sqrt{2}$ for
        the two-state and three-state models, respectively.  {Note that the PL curves do not exist above $\rho^*=1$.}
  \label{fig:EOS_etc_dumbbell}}
\end{figure}

     This
     shortcoming of the PL theory and the deviation from additive prism
     systems is illustrated in Fig.~\ref{fig:EOS_etc_dumbbell}, where
     the exact and the PL results are shown together for the bulk
     properties of 2- and three-state hard dumbbells.

     In the two-state model, only  the contact distance $\sigma_{13}\simeq 0.91\times \left(\sigma_{1}+\sigma_{3}\right) / 2$ induces
     some negative non-additive effects in the results. Therefore,  due
     to the very weak non-additive character of $\sigma_{13}$,
     the resulting equation of state and the ordering properties are
     almost identical with those of two-state hard prisms. Similarly to hard
     prisms, two-state hard dumbbells undergo a continuous structural change
     from the isotropic fluid to the perfectly ordered nematic one
     with increasing density. This can be seen clearly in
     Fig.~\ref{fig:EOS_etc_dumbbell}(b), where the fraction of
     particles in state $1$ ($x_{1}$) increases continuously
     from $0.5$ to $1$. In the two-state dumbbell model, $\eta$ cannot be
     considered as a 1D packing fraction, but it becomes identical
     with the packing fraction of hard rods of length $\sigma$ at
     high densities ($\rho^{*}>0.8$). It can be seen in
     Fig.~\ref{fig:EOS_etc_dumbbell}(c) that $\eta$ is higher than the
     packing fraction of hard rods at intermediate densities since
     there are some dumbbells with  length $2 \sigma$ along the $z$
     axis. This positive deviation is due to the orientational
     entropy, which favors the orientational disorder. However, this
     entropy term weakens with increasing density due to the
     decreasing available room.

     To analyze the deviation from the additive hard-prism system, we plot $(1-\eta)\beta P / \rho$, which must be equal to $1$ for additive systems [see
     Eq.~(\ref{eq:P_additive})], as a function of
     density in Fig.~\ref{fig:EOS_etc_dumbbell}(d). We can see that the two-state dumbbell
     fluid produces  values lower than $1$ at intermediate densities since two dumbbells can get closer to each other in
     perpendicular orientation than two prisms can,
     i.e., $\sigma_{13}<\left(\sigma_{1}+\sigma_{3}\right) / 2$, which
     manifests in a lower pressure and a higher packing fraction at a
     given density. Figure~\ref{fig:EOS_etc_dumbbell}(d) shows that the
     PL theory underestimates the effect of non-additivity,  producing higher values for $(1-\eta)\beta P / \rho$ than the exact results obtained from the
     $\mathrm{TM}$ and ND methods. Apart from this deviation, the
     PL theory describes accurately all quantities of the weakly
     non-additive system of two-state dumbbells.

     The phase behavior of the three-state hard dumbbell system is more
     complicated due to the inclusion of state 2, which is a
     competitor of state 1 in the ordering process. This can be seen
     in Fig.~\ref{fig:EOS_etc_dumbbell}, where the range of
     dimensionless density ($\rho^{*}$) extends to
     $\sqrt{2}$ due to the extra orientation state. Therefore, the
     equation of state of the three-state system deviates substantially from
     that of the two-state one, as can be seen in
     Fig.~\ref{fig:EOS_etc_dumbbell}(a). We observe that the three-state
     pressure curve is below the two-state one because there is more
     space between particles with the inclusion of state 2. This is
     due to the orientational entropy, which is maximal if the number
     of particles is the same in all orientations. This entropy term
     prevails at very low densities, where the system has an isotropic
     distribution, i.e., $x_{1}=x_{2}\simeq x_{3}\simeq 1 / 3$. However, the
     competition between different entropy terms produces
     orientational ordering in such a way that $x_{1}=x_{2} \to
     1/2$ and $x_{3} \rightarrow 0$ at close packing, as clearly observed in Fig.~\ref{fig:EOS_etc_dumbbell}(b).  The orientationally ordered structure
     develops at $\rho^{*} \approx 1$, where the average distance
     between the neighboring particles reduces to $\sigma$, which do
     not allow the particles to occupy a distance $2 \sigma$  along the $z$
     axis, i.e., $x_{3} \approx 0$.

     One might think naively
     that the phase behavior of three-state dumbbells becomes identical
     with that of hard rods of length $\sigma$ for $\rho^*>1$ because particles in states
     1 and 2 occupy the same distance along the $z$ axis. This idea would be
     supported by Fig.~\ref{fig:EOS_etc_dumbbell}(c), where the curve $\eta$
     vs $\rho^*$  becomes linear for $\rho^*>1$, as in the fluid of
     hard rods. However, this equivalence turns out not to be true, since the entropic
     contributions of $\sigma_{11}=\sigma_{22}$ and $\sigma_{12}$ contact
     distances are still dominant for $\rho^{*}>1$. The difference
     between the three-state hard dumbbell and  additive hard-body
     fluids can be visualized with the help of $(1-\eta)\beta P/\rho$,
     which is shown as a function of density in
     Fig.~\ref{fig:EOS_etc_dumbbell}(d). We can see that the three-state
     hard dumbbells do not obey $(1-\eta)\beta P/\rho=1$ because the
     non-additivity decreases pressure and increases $\eta$ at a given
     density, as compared to an additive system. Moreover, it
     changes sign at $\eta=1$ (which corresponds to $\rho^*\simeq 1$). Therefore, the equation of state of
     three-state dumbbells cannot be mapped onto that of hard rods of length $\sigma$, even at
     very high densities. Instead of random orientational ordering of
     states 1 and 2 along the $z$ axis, the particles form clusters,
     where neighboring particles are perpendicular to each other,
     i.e., dimers, trimers, tetramers, \ldots, $m$-mers form with
     increasing density. At close packing, the length of the cluster
     must go to infinity and the structure is crossed through the
     whole system to reach the maximal density. As a consequence, what actually happens is that the three-state dumbbell model for $\rho^*>1$ becomes progressively closer to  hard rods of length $\sigma_{12}=\sigma/\sqrt{2}$, so that $\beta P/\rho\to (1-\rho^*/ \sqrt{2})^{-1}$ as density approaches its close-packing value $\rho_{\text{cp}}^*=\sqrt{2}$.

     Regarding the PL
     theory as applied to the three-state model, it is accurate for the ordering properties [see
     Fig.~\ref{fig:EOS_etc_dumbbell}(b)], but it fails to predict
     the crossed structure. This can be seen in the equation of state
     [see Figs.~\ref{fig:EOS_etc_dumbbell}(a) and \ref{fig:EOS_etc_dumbbell}(d)], in which case the pressure of the
     PL theory diverges at $\rho^*=1$, precisely where the crossed structure
     starts to develop.

     \subsection{Pair distribution function}
We can get more insight into the ordering properties of hard
     prisms and hard dumbbells by studying the positional
     pair distribution function  along the channel, $g_{ij}(z)$.
     \subsubsection{Hard prisms}

\begin{figure}
      \includegraphics[width=\columnwidth]{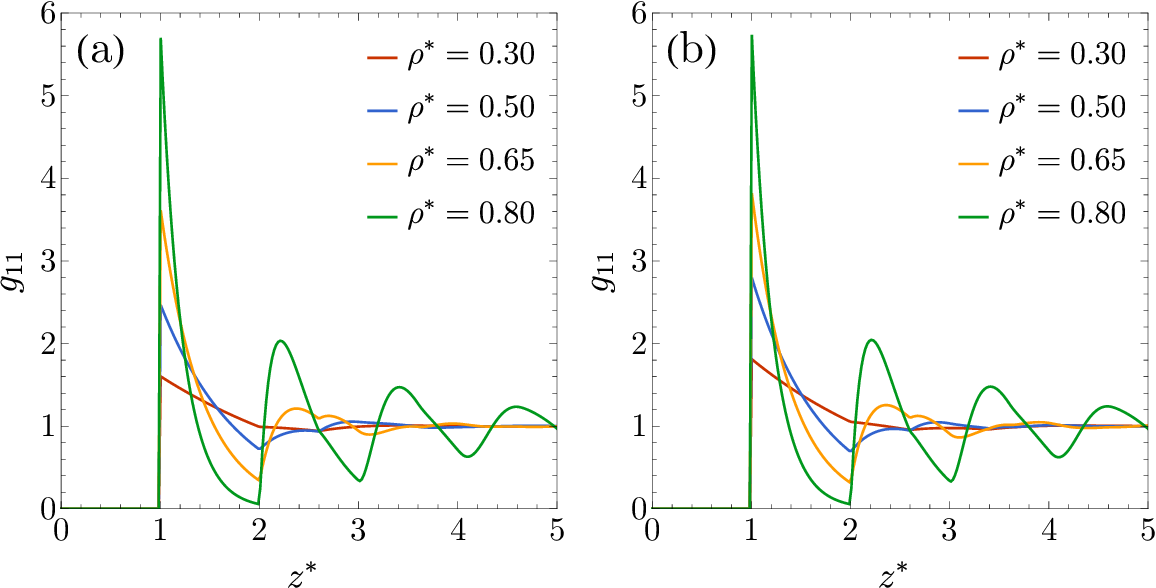}
      \caption{Pair distribution function between two hard
        prisms, both having a  length $\sigma_{1}$ along the $z$ axis, as a
        function of distance between the two particles
        ($z^{*}$). The results are shown for (a) two-state
        and (b) three-state systems for densities $\rho^{*}=0.3$, $0.5$, $0.65$,
        and $0.8$. The other pair distribution functions can be
        obtained from $g_{11}$ by applying the shift property, Eq.~\eqref{eq:shift}.
  \label{fig:g_prism}}
\end{figure}

      As the hard  prisms obey the shift property of additive $1 \mathrm{D}$ hard-body mixtures [see Eq.~\eqref{eq:shift}],\cite{Grodon_2004,Grodon_2005,Schmidt_2007} only
     $g_{11}(z)$ is shown in Fig.~\ref{fig:g_prism}.
          We can see that the prisms become more and more ordered locally
     and the positional order propagates to larger and larger
     distances with increasing density. At low densities, $g_{11}(z)$
     is structureless because the orientational ordering is very
     weak, while it becomes oscillatory at high density with
     the period of shortest length $\left(\sigma_{1}\right)$ due to
     the development of perfect nematic order with
     $x_{1}=1$. Therefore, $g_{11}(z)$ becomes identical to $g(z)$ of
     hard rods having a diameter $\sigma_{1}$  at very high
     densities. On the other hand, $g_{11}(z)$ is more structured at
     intermediate densities ($\rho^{*}=0.5$ and $\rho^{*}=0.65$),
     where the fractions of particles having lengths $\sigma_{2}$ and
     $\sigma_{3}$  along the $z$ axis are not negligible. The
     effects of $x_{2}$ and $x_{3}$ on $g_{11}(z)$ are present in new
     singularities at $z=i \sigma_{1}+j \sigma_{2}+k \sigma_{3}$,
     where $i, j, k$ are positive integers [see Eq.~\eqref{3.29}]. Among those singularities,  three of them are kinks at $z=2\sigma_1$, $\sigma_1+\sigma_2$, and $\sigma_1+\sigma_3$ (the latter only in the three-state model), which correspond to $\ell=2$ in Eqs.~\eqref{eq:qikj} and \eqref{3.29}; the other singularities are of higher order and thus they are not visible in Fig.~\ref{fig:g_prism}. This
     shows that $g_{i j}(z)$ cannot be mapped onto an effective hard-rod $g(z)$ with $d=\langle \sigma\rangle$,
     since in the latter the singularities appear only at multiples of $d$.

     We can also see in Fig.~\ref{fig:g_prism} that the pair distribution functions of the two-state and three-state prisms at a common density are almost identical, the main difference being that the
     three-state prisms are positionally slightly more ordered than the two-state ones at intermediate densities
     because the three-state system has a higher packing fraction [see
     Fig.~\ref{EOS_etc_prism}(c)].

     \subsubsection{Hard dumbbells}

     Now we turn our attention to the
     positional ordering of hard dumbbells. Since the shift property
     of additive fluids is not valid for hard dumbbells, $g_{11}(z)$,
     $g_{13}(z)$, and $g_{33}(z)$ are calculated for two-state dumbbells,
     whereas $g_{11}(z)=g_{22}(z)$, $g_{12}(z)$, $g_{13}(z)=g_{23}(z)$, and $g_{33}(z)$ are considered for the
     three-state model.
\begin{figure}
      \includegraphics[width=\columnwidth]{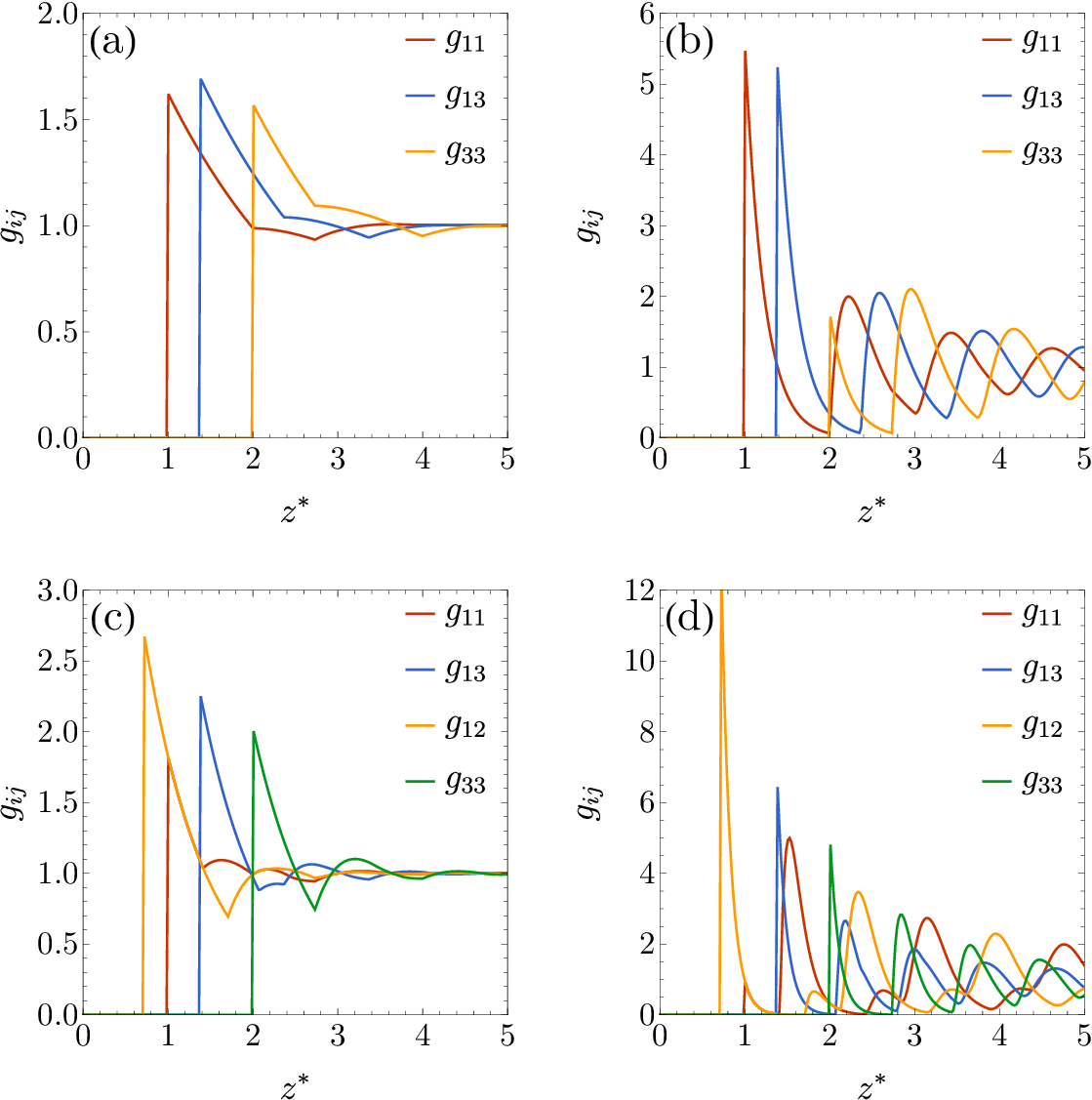}
      \caption{Pair distribution function between two hard dumbbells,
        having lengths $\sigma_{i}$ and $\sigma_{j}$
        along the $z$ axis, as a function of distance between the two
        particles ($z^{*}$). The results are shown for (a) the
        2 -state system at $\rho^{*}=0.3$, {(b)} the two-state system at $\rho^*=0.8$, (c) the
        three-state system at $\rho^{*}=0.6$, and  (d)  the
        three-state system at $\rho^{*}=1.2$.
  \label{fig:g_dumbbell}}
\end{figure}

     We can see that the shift property of $g_{i j}(z)$ obtained for
     additive systems is violated even at $\rho^{*}=0.3$ in the
     two-state model, although the shapes of all $g_{i j}(z)$ are very
     similar [see Fig.~\ref{fig:g_dumbbell}(a)]. However, the pair distribution functions
     become very different from each other at high densities [see
     Fig.~\ref{fig:g_dumbbell}(b)], since the dumbbells tend to align with
     their short lengths ($\sigma$) along the $z$ axis. The
     consequence of this fact is that $g_{11}(z)$ becomes very similar
     to the pair distribution function $g(z)$ of hard rods having a diameter $\sigma$, whereas
     $g_{33}(z)$ shows a second peak higher than the first one at
     $\rho^{*}=0.8$. This peculiar behavior of $g_{33}(z)$ is due to
     the fact that two dumbbells having  lengths $2\sigma$ do not like
     to form a neighboring pair. Opposite to this, $g_{13}(z)$ is very similar to
     $g_{11}(z)$ because two particles like to form a pair if they
     have different orientations. In fact, it can be proved from Eq.~\eqref{3.5cc} for the two-state model that, in the high-density limit, the shift property {$g_{13}(\sigma_{13}+ z)\simeq g_{11}(\sigma_1+ z)$} is fulfilled. Also in that limit, $g_{33}(z)$ depletes in the interval $\sigma_3<z<2\sigma_{13}$. Beyond $z=2\sigma_{13}$, $g_{33}(z)$ replicates the behavior of $g_{11}(z)$ for $z>2\sigma_1$, i.e., {$g_{33}(2\sigma_{13}+ z)\simeq g_{11}(2\sigma_1+ z)$}.

     Figures~\ref{fig:g_dumbbell}(c) and \ref{fig:g_dumbbell}(d) show that  three-state
     hard dumbbells behave very differently because the shortest
     distance between two dumbbells reduces to $\sigma / \sqrt{2}$. At
     this distance the neighboring particles are perpendicular to
     each other, and $g_{12}$ has the highest contact value among all
     $g_{i j}$. Due to the favorable crossed alignments, the pair distribution functions of the three-state system are more structured than that of the two-state
     one at both low and high densities. In addition to this, the
     distance between peaks of $g_{i j}$ is shorter in the three-state
     model, since the particles can get closer to each other in
     crossed ordering. We can see in Fig.~\ref{fig:g_dumbbell}(d) that the structures of $g_{11}$ and
     $g_{12}$ are very special at the high density $\rho^{*}=1.2$, because the first,
     third, fifth, \ldots\ peaks of $g_{11}$ ($g_{12}$) show
     increasing (decreasing) trends, while the second, fourth, sixth,
     \ldots\ peaks  exhibit the opposite trends. Moreover, the second peak of
     $g_{11}$ ($g_{12}$) is much higher (smaller) than the
     first one. Therefore, the trends observed in $g_{11}$ and $g_{12}$
     prove that the first neighbors like to be perpendicular, while
     the second ones tend to be parallel to each other, i.e., particles form
     crossed clusters in 1-2-1-2-$\cdots$ orientational order. We
     mention that  the distribution functions $g_{13}$ and $g_{33}$ do not
     provide information about the crossed order, but they show
     enhanced positional order at $\rho^{*}=1.2$.

     \subsection{Correlation lengths}
\begin{figure}
      \includegraphics[width=\columnwidth]{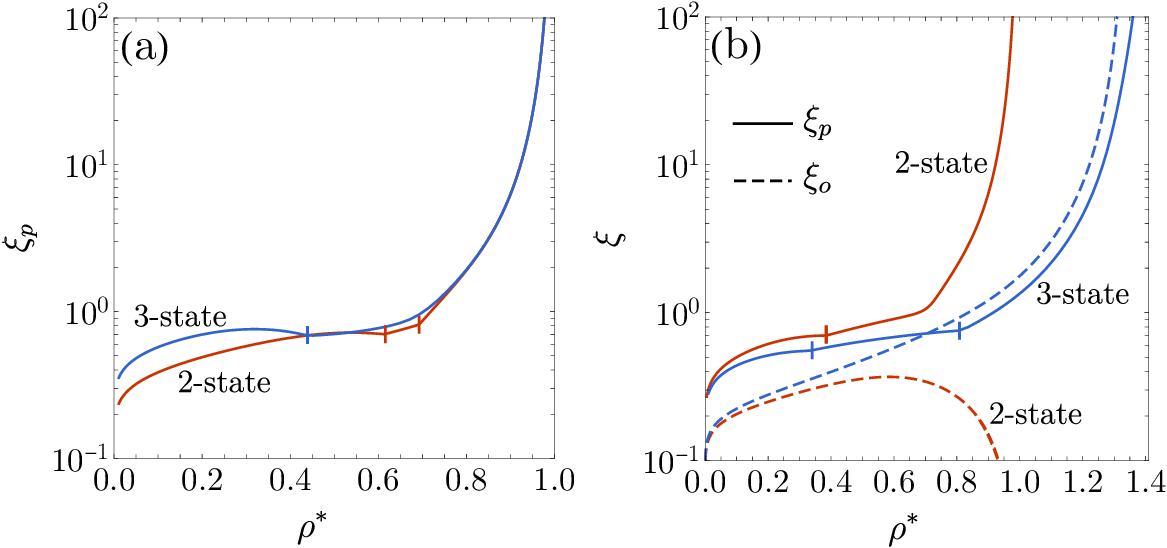}
      \caption{Orientational ($\xi_{o}$) and positional ($\xi_{p}$)
        correlation lengths as  functions of density. Results for hard
        prisms are shown in panel (a), while those for hard dumbbells are shown in panel
        (b). The vertical bars show  the location of kinks apparent in
        $\xi_{p}$.
  \label{fig:corr_length}}
\end{figure}

     The extent of
     positional order and the propagation of orientational ordering
     are measured with the help of positional and orientational
     correlations lengths, which are shown as a function of density in
     Fig.~\ref{fig:corr_length}.

\subsubsection{Hard prisms}

     The hard prisms are not  orientationally correlated because only
     the highest eigenvalue of the transfer matrix is nonzero, while
     the other ones are zero. This results in $\xi_{o}=0$ for both 2- and
     three-state prisms. The positional correlation of hard prisms
     ($\xi_{p}$) coming from the  oscillatory exponential decay of the pair distribution functions [see Eq.~\eqref{hij}] has quite a complicate
     dependence on  density  because of the crossing of the real part of two poles with different imaginary parts (structural crossover). As a consequence, two kinks are present in the     two-state prism model, while only one in the three-state one. This can be
     seen in Fig.~\ref{fig:corr_length}(a), where the nonmonotonic
     behavior of $\xi_{p}$ is shown for both prism models. The
     presence of kinks and the saturation of $\xi_{p}$ at intermediate
     densities may be the result of competition between orientational
     and positional ordering, because the orientational ordering
     weakens the positional correlations and shifts the positional
     ordering towards higher densities. It can also be seen that the
     three-state system is more correlated than the two-state one at
     intermediate densities. This may be due to the fact that the three-state
     system is more packed than the two-state one at a given
     density, as shown in Fig.~\ref{EOS_etc_prism}(c). For densities $\rho^*>0.8$, both models become equivalent to a 1D hard-rod fluid and, thus, they have the same positional correlation length.

\subsubsection{Hard dumbbells}

     Figure~\ref{fig:corr_length}(b) shows that the
     positional correlation of hard dumbbells is qualitatively similar to that of
     hard prisms, with the difference that the two-state dumbbell system
     has only one kink, while two kinks are present in the three-state
     model. Moreover, at the same $\rho^*$, the positional correlation is stronger in the
     two-state model because it is more packed at a given density [see Fig.~\ref{fig:EOS_etc_dumbbell}(c)]. The
     result of packing effects is that the positional correlation length of
     hard dumbbells diverges at $\rho^{*}=1$ in the two-state model,
     while this happens at $\rho^{*}=\sqrt{2}$ in the three-state one. The
     orientational correlation is very weak and only present at
     intermediate densities in the two-state system. This is due to the
     fact that the dumbbells with  length $2\sigma$ like to form
     pairs with dumbbells with  length $\sigma$, but the fraction of the former
     dumbbells  decreases with density. This is
     not the case in the three-state model, where the orientational
     correlation length diverges at $\rho^{*}=\sqrt{2}$.

     Therefore, three-state dumbbells exhibit long-range orientational
     and positional correlation near close packing. This is
     the consequence of the crossed close-packing
     structure, where the dumbbells form infinitely long
     clusters with  an orientation sequence 1-2-1-2$\cdots$. We believe
     that these findings keep being true even in the freely rotating case
     because the close-packing structure does not change.

     \section{Conclusions}\label{sec5}

     In this paper, we have examined the effect of additive and non-additive hard-body
     interactions on the phase behavior of q1D hard-body fluids,
     where the particles are allowed to move freely along a straight
     line and to rotate into a finite number ($n$) of orientational
     states. Only two perpendicular orientations are allowed in the
     two-state model ($n=2$), while three mutually perpendicular ones are present in
     the three-state model ($n=3$). The two-state model can be considered as
     a minimal model of some single-file fluids placed in a nanopore
     with rectangular cross section, while the three-state one can
     represent a single-file fluid in a cylindrical pore.

     The additive
     system has the feature that the contact distance between two
     particles obeys $\sigma_{i j}=\left(\sigma_{i}+\sigma_{j}\right)
     / 2$, where $\sigma_{i}$ and $\sigma_{j}$ are the lengths of a
     particle with orientations $i$ and $j$ along the $z$ axis, respectively. The
     prototypes of additive systems are hard spheres, but
     prisms can also be additive in the above 2- and three-state
     representation. However, some systems can deviate into positive
     $\left(\sigma_{i j} \geq\left(\sigma_{i}+\sigma_{j}\right) /
     2\right)$ or negative $\left(\sigma_{i j}
     \leq\left(\sigma_{i}+\sigma_{j}\right) / 2\right)$ direction from
     the additive systems. In this regard, the hard-dumbbell model belongs
     to the class of negative non-additive systems.

     We found that the phase
     behaviors of additive and non-additive systems differ
     significantly. While the equation of state of additive systems
     can be mapped onto that of 1D hard rods with an effective length $d=\langle\sigma\rangle$, this is not so for the non-additive systems. We found that $(1-\eta)\beta P / \rho$ can measure
     the effect of non-additive interaction since that quantity is exactly equal to $1$ for all $n$-state additive systems. By including just
     one non-additive contact distance, as in the two-state
     dumbbell model, $(1-\eta)\beta P / \rho$ deviates from $1$ only
     at intermediate densities, since the two-state dumbbell system
     becomes identical to the fluid of hard rods as it approaches
     close packing. This comes automatically from the $\mathrm{TM}$
     method, where $K_{11}=e^{-\beta P \sigma_{1}} / \beta P$ becomes
     the dominant {matrix element} and determines the phase behavior at high
     pressures, assuming $\sigma_{1}=\sigma$ is the shortest contact
     distance. In the three-state model, the deviation from the 1D hard-rod behavior is much more pronounced
      due to the presence of $\sigma_{12}=\sigma /
     \sqrt{2}$. The consequence of this
     non-additive interaction is that the particles like to form
     crossed clusters and $(1-\eta)\beta P / \rho$ becomes negative
     for $\rho^{*}>1$. In the three-state dumbbell model, the dominant
     transfer matrix element is $K_{12}=e^{-\beta P \sigma_{12}} /
     \beta P$, while the other elements can be neglected at high
     pressures. It can be  easily shown in this limit that the largest and second largest (in absolute) value eigenvalues are
     $\lambda\to K_{12}$ and  $\lambda^{*}\to-K_{12}$, respectively. Therefore, the
     resulting equation of state of the system at high pressures is
     given by $\beta P=\rho /(1-\rho \sigma / \sqrt{2})$, which
     corresponds to an equation of state of $1 \mathrm{D}$ hard rods
     having a length $\sigma / \sqrt{2}$. However, the structure is
     crossed at high pressures and the orientational correlation
     length diverges since
     $\xi_{o}^{-1}=\ln \left(\lambda/|\lambda^{*}|\right)\to 0$. This
     argument is strictly valid only in the limit $\beta P \rightarrow \infty$,
     which corresponds to the close-packing density, while the size of
     the crossed cluster is finite for densities below the close-packing one.

     We showed that the general additive q1D fluids can be studied exactly
     using the PL theory, as well as the TM and ND methods. While the PL theory provides
     only the bulk properties, such as the equation of state and the
     orientation order parameter, the TM and ND methods can also be used to
     determine the local structure of the systems. The study of the
     structural and bulk properties of q1D hard dumbbells revealed the
     importance of non-additive interactions, which can be the driving
     force in the formation of complex necklace-like structures of
     anisotropic building blocks, such as the crossed structure. Those
     systems can be studied exactly using the TM and ND methods, but
     the PL theory cannot account for the high-density and close-packing structures. Therefore, the success of the PL theory of
     anisotropic 2D and 3D hard-body fluids may be due to the correct
     description of side-by-side and end-to-end configurations even if
     the effect of ``T'' and other intermediate configurations are
     incorrectly included into the theory. Regarding the role of the
     DFT, the exact density functional of additive 1D hard-body
     $n$-component mixtures, which was devised by
     Vanderlick et al.,\cite{Vanderlick_1989} can be applied to one-component anisotropic hard-body fluids with $n$ orientational
     states using the equal chemical potential
     condition.\cite{Onsager_AnnNYAcadSci_1949} It can be shown that
     hard prisms can be described exactly within the DFT, while
     hard dumbbells cannot. It is worth noting that  the ND method
     pointed out the weaknesses of the DFT in describing the
     positive and negative non-additive 1D hard-body
     mixtures.\cite{Santos_2007} Therefore, the only possible way to
     get exact results for more realistic systems, such as the freely
     rotating q1D rods and single-file fluids in cylindrical pore, is
     to use the TM and the ND methods. We believe that these exact
     methods can explain some of the simulation results on the
     ordering properties of real rod-like nanoparticles, which can
     move freely in q1D diblock copolymer templates.\cite{Tang_2009}

     Finally, it must be stressed that the results presented in Sec.~\ref{sec3} are not restricted to the specific 2- and three-state prism and dumbbell models, chosen here as prototypes of additive and non-additive systems, respectively. In the case of additive interactions, the high-density phase is equivalent to that of a monocomponent fluid with the smallest length. Interestingly, the same situation occurs if the non-additivity is positive [i.e., $\sigma_{ij}\geq (\sigma_i+\sigma_j)/2$] or if it is negative but the smallest length  is smaller than any cross distance $\sigma_{ij}$, as happens with the two-state dumbbell model studied in this paper. However, if the smallest cross distance is smaller than the smallest length, then the high-density phase presents the crossed structural ordering exemplified here by the three-state dumbbell model. In this respect, the two-state hard dumbbell model is a caricature version of a model where the dumbbells can freely rotate on the $xz$ plane in a rectangular channel; analogously, if orientation 3 is removed from our three-state hard dumbbell model, one has a simplification of a more general model in which the dumbbells can freely rotate on the $xy$ plane in a circular channel. Preliminary results show that the orientational and positional correlation lengths of the continuous models are qualitatively similar to those of the discrete models investigated in this paper. Work is currently in progress to study these cases with continuous orientations and the results will be published elsewhere.

\acknowledgments
A.M.M. and A.S. acknowledge financial support from Grant No.~PID2020-112936GB-I00 funded by MCIN/AEI/10.13039/501100011033 and from Grant No.~IB20079 funded by Junta de Extremadura (Spain) and by ``ERDF A way of  making Europe.''
A.M.M. is also grateful to MCIN/AEI/10.13039/501100011033 and  ``ESF Investing in your future'' for a predoctoral fellowship PRE2021-097702.
S.V. and P.G. gratefully acknowledge the financial support of the National
Research, Development, and Innovation Office - Grant
No.~NKFIH K137720 and TKP2021-NKTA-21.

\section*{AUTHOR DECLARATIONS}
\subsection*{Conflict of Interest}
The authors have no conflicts to disclose.
\subsection*{Author Contributions}

\textbf{Ana M. Montero}: Formal analysis (supporting); Investigation (equal);
Methodology (supporting); Software (lead); Visualization (lead).
\textbf{Andr\'es Santos}: Conceptualization (equal); Formal analysis
(equal); Funding acquisition (equal); Investigation (equal); Methodology
(supporting); Supervision (equal);  Writing -- original draft (supporting);
Writing -- review \& editing ({lead}).
\textbf{P\'eter Gurin}: Conceptualization (equal);  Formal analysis
(equal); Funding acquisition (equal); Investigation (equal); Methodology (lead); Validation (equal); Writing -- original draft (supporting); Writing -- review \& editing (supporting).
\textbf{Szabolcs Varga}: Conceptualization (lead);  Formal analysis
(equal); Funding acquisition (equal); Investigation (equal);  Methodology (equal); Validation (equal); Writing -- original draft (lead); Writing -- review \& editing (supporting).

\section*{Data availability}
The data that support the findings of this study are available from the corresponding author upon reasonable request.

\bibliography{ordering_of_anisotropic_particles_in_channel}
\end{document}